\documentclass[sigconf]{acmart}
\setcopyright{none}
\usepackage{amsmath,amsfonts,bm}

\def\eqref#1{equation~\ref{#1}}

\def\1{\bm{1}}

\DeclareMathAlphabet{\mathsfit}{\encodingdefault}{\sfdefault}{m}{sl}
\SetMathAlphabet{\mathsfit}{bold}{\encodingdefault}{\sfdefault}{bx}{n}

\usepackage{microtype}
\usepackage{graphicx}
\usepackage{booktabs}
\usepackage{tabularx}
\usepackage{multirow}
\usepackage[ruled,vlined,linesnumbered]{algorithm2e}
\DontPrintSemicolon
\usepackage{booktabs}
\usepackage[table]{xcolor}
\definecolor{scoutscore}{RGB}{255,244,230} 
\definecolor{scoutadapt}{RGB}{236,247,237} 
\usepackage{placeins} 

\title{A Practical Framework for Flaky Failure Triage in Distributed Database Continuous Integration}

\author{Jun-Peng Zhu}
\authornote{These authors contributed equally to this work.}
\affiliation{
  \institution{Northwest A\&F University PingCAP}
  \country{China}
}
\email{zjp.dase@nwafu.edu.cn}

\author{Qizhi Wang}
\authornotemark[1]
\authornote{Corresponding author.}
\affiliation{
  \institution{PingCAP}
  \country{China}
}
\email{qizhi.wang@pingcap.com}

\author{Yulong Zhai}
\affiliation{
  \institution{PingCAP}
  \country{China}
}
\email{zhaiyl@pingcap.com}

\author{Yishen Sun}
\affiliation{
  \institution{PingCAP}
  \country{China}
}
\email{sunyishen@pingcap.com}

\author{Sen Chen}
\authornotemark[2]
\affiliation{
  \institution{Nankai University}
  \country{China}
}
\email{senchen@nankai.edu.cn}

\author{Xu Kai}
\affiliation{
  \institution{PingCAP}
  \country{China}
}
\email{xukai@pingcap.com}

\author{Peng Cai}
\affiliation{
  \institution{East China Normal University}
  \country{China}
}
\email{pcai@dase.ecnu.edu.cn}

\author{Hongming Zhang}
\affiliation{
  \institution{Northwest A\&F University}
  \country{China}
}
\email{zhm@nwsuaf.edu.cn}

\author{Heng Long}
\affiliation{
  \institution{PingCAP}
  \country{China}
}
\email{lh@pingcap.com}

\author{Liu Tang}
\affiliation{
  \institution{PingCAP}
  \country{China}
}
\email{tl@pingcap.com}

\author{Qi Liu}
\affiliation{
  \institution{PingCAP}
  \country{China}
}
\email{liuqi@pingcap.com}

\begin{document}
\begin{abstract}
    Flaky failure triage is crucial for keeping distributed database continuous integration (CI) efficient and reliable. After a failure is observed, operators must quickly decide whether to auto-rerun the job as likely flaky or escalate it as likely persistent, often under CPU-only millisecond budgets. Existing approaches remain difficult to deploy in this setting because they may rely on post-failure artifacts, produce poorly calibrated scores under telemetry and workload shifts, or learn from labels generated by finite rerun policies. To address these challenges, we present SCOUT, a practical  \textbf{S}tate-aware \textbf{C}ausal \textbf{O}nline \textbf{U}ncertainty-calibrated \textbf{T}riage framework for distributed database CI. SCOUT uses only strict-causal features, including pre-failure telemetry and strictly historical data, to make online decisions without lookahead. Specifically, SCOUT combines lightweight state-aware scoring with optional sparse metadata fusion, applies post-hoc calibration to support fixed-threshold decisions across temporal and cross-domain shifts, and introduces a posterior-soft correction to reduce label bias induced by finite rerun budgets. We evaluated SCOUT on a benchmark of 3,680 labeled failed runs, including 462 flaky positives, and 62 telemetry/context features. Further, we studied the feasibility of SCOUT on TiDB v7/v8 and a large GitHub Actions metadata-only trace. The experimental results demonstrated its effectiveness and usefulness. We deployed SCOUT in the production environment, achieving an end-to-end P95 latency of 1.17 ms on CPU.
\end{abstract}


\maketitle

\section{Introduction}
Continuous integration (CI) is a core mechanism for maintaining software quality, but in distributed database systems, CI failures are often linked to transient runtime state rather than to persistent defects alone. Lock contention spikes, replication lag, storage stalls, and queue bursts can cause a run to fail even when the underlying code or test is not consistently faulty. As a result, once a failure is observed, operators must quickly decide whether to auto-rerun the job as likely flaky or escalate it as likely persistent. This decision is operationally important because unnecessary escalation wastes debugging effort, while unnecessary reruns delay diagnosis and consume scarce CI resources.

Prior work on flaky tests and CI failures has made important progress, but it does not directly solve this deployment setting. Some approaches study flaky test prediction using code evolution and test history~\cite{historychurn2023, chromium2023}, while others detect or categorize failures from post-failure symptoms and logs~\cite{jitmatch2023}. These works are valuable, but they address different pain points in the CI lifecycle. The history-based methods are more naturally used before execution, whereas symptom-matching methods typically depend on post-failure artifacts. Meanwhile, systems and AIOps research increasingly use multimodal logs and metrics for diagnosis and operations management~\cite{liveforensics2019,agentfm2025,logdb2025}. However, such approaches are generally aimed at rich diagnosis or root-cause analysis rather than an immediate rerun-versus-escalate decision under strict latency and resource~constraints.

This gap matters because online flaky-failure triage is not simply a matter of training a larger model to recognize flakiness. Instead, it raises three deployment challenges. \textbf{(C1) Leakage-sensitive features}. Feature leakage is easy to introduce. Post-failure logs, error strings, or future-run information can improve apparent predictive performance while violating the online decision protocol. \textbf{(C2) Threshold-dependent decisions}. The downstream action depends on a thresholded decision, not merely on ranking quality. Post-hoc calibration and calibration under shift are well studied~\cite{niculescu2005,pampari2020}, and recent work further emphasizes calibration as a decision primitive under explicit costs~\cite{elkan2001foundations, gopalan2025efficientcalibdecisions}. In our setting, this means that a model with acceptable PR-AUC can still behave poorly if its scores are miscalibrated under temporal or cross-domain shifts. \textbf{(C3) Policy-generated labels}. Rerun-based flakiness labels are policy-generated. With a finite rerun budget, some flaky failures are systematically under-observed, making the problem closely related to positive-unlabeled learning under selection bias~\cite{bekker2018sarem,elkan2008pu} and to broader censored-outcome settings studied in econometrics and survival analysis~\cite{kaplan1958,tobin1958,cox1972}. Modern selective-prediction and uncertainty-control methods~\cite{tibshirani2019conformalshift, vennabers2012, chow1970reject, weightedconformal2024, angelopoulos2021conformal} are also relevant here, since real deployments must often hedge when calibration data are scarce, or target overlap is poor.

To address these challenges, we present SCOUT, a practical \textbf{S}tate-aware \textbf{C}ausal \textbf{O}nline \textbf{U}ncertainty-calibrated \textbf{T}riage framework for distributed-database CI. SCOUT addresses \textbf{C1} by using only strict-causal inputs, including pre-failure telemetry and strictly historical data, to make online rerun-versus-escalate decisions without lookahead. It addresses \textbf{C2} through lightweight state-aware scoring with optional sparse metadata fusion and post-hoc calibration, enabling a single cost-derived threshold to transfer across shifts. It addresses \textbf{C3} through a posterior-soft correction that reduces the label bias induced by finite rerun budgets.

We evaluate SCOUT on a benchmark of 3,680 labeled failed runs, including 462 flaky positives, and use 62 telemetry/context features.
We also evaluate the feasibility of TiDB v7/v8 and a large GitHub Actions metadata-only trace.
The results show that low-level runtime telemetry carries most of the actionable signal in this setting, while calibrated probabilities matter more than heavier model families for portable actions. Under temporal shift, isotonic calibration reduces fixed-threshold decision cost from 765.4 to 496.9. Posterior-soft correction substantially improves calibration to the larger-budget oracle label, reducing ECE from 0.320 to 0.027 and Brier score from 0.213 to 0.096, while remaining robust under correlated reruns.
SCOUT also satisfies practical deployment constraints, achieving an end-to-end P95 latency of 1.17 ms on CPU.

To summarize, this paper makes the following contributions:

(1) We formulate flaky-failure triage in distributed database CI as a strict-causal online decision problem, where the system must decide whether to auto-rerun a failed run as likely flaky or escalate it as likely persistent under CPU-only millisecond budgets.

(2) We propose \textsc{SCOUT}, a practical framework for this setting that combines strict-causal feature extraction, lightweight state-aware scoring, and thresholded online decision-making using only pre-failure telemetry, governed pre-run metadata, and strictly~earlier-run~history.

(3) We develop two deployment-oriented techniques for reliable triage: decision-portable calibration for fixed-threshold actions under temporal and cross-domain shift and posterior-soft rerun-budget correction for mitigating the label bias induced by finite rerun policies.

(4) We conduct a comprehensive evaluation on synthetic benchmarks, a real TiDB v7/v8 trace, and a large GitHub Actions metadata-only trace, and further demonstrate that SCOUT is practical in production, achieving an end-to-end P95 latency of 1.17 ms on~CPU.

\section{Preliminaries}

In this section, we introduce the preliminaries of this paper.

\textit{\textbf{Online flaky-failure triage}}. Consider a failed primary CI run $i$. Let $x_i$ denote the evidence available at decision time, and let $y_i \in \{0,1\}$ denote whether the observed failure is flaky, where $y_i=1$ means that the failure is non-reproducible under reruns and $y_i=0$ means that it is persistent. A triage policy outputs a score
\[
p_i \approx P(y_i = 1 \mid x_i),
\]
and maps this score to an action
\[
a_i \in \{\texttt{rerun}, \texttt{escalate}\}.
\]
The action \texttt{rerun} means that the failed job is automatically retried, while \texttt{escalate} means that the failure is surfaced for debugging or downstream diagnosis.

\textit{\textbf{Strict-causal features}}.
Each run has a telemetry window around the failure timestamp. Our main benchmark uses only the pre-failure window $[-120\text{s}, 0]$. We treat $[-120\text{s}, +30\text{s}]$ as an explicit leakage ablation (post-failure evidence).

A CI failure triggers strict-causal feature extraction from pre-failure telemetry and history. Lightweight models produce a score, which is optionally calibrated for decision portability and then mapped to an action via a fixed threshold. Labels are generated by a finite rerun policy, motivating rerun-budget correction.

We aggregate only pre-failure telemetry $[-120\text{s}, 0]$ into dense features and use history computed only from earlier primary runs. Post-failure artifacts (including error strings/logs) are blocked and used only in explicit leakage ablations. Also used are pre-run metadata tokens.

\textit{\textbf{Rerun labels and finite-budget semantics}}.
In practice, flaky-failure labels are generated by a finite rerun policy rather than observed directly. For a failed primary run, let $R$ be the available rerun budget. We define the budget-$R$ label as
\[
y_i^{(R)} =
\begin{cases}
1, & \text{if at least one rerun passes within } R \text{ attempts},\\
0, & \text{otherwise}.
\end{cases}
\]
Under this definition, a failure is labeled flaky if it becomes non-reproducible within the available rerun budget. The label is monotone in the budget:
\[
R_1 \le R_2 \;\Longrightarrow\; y_i^{(R_1)} \le y_i^{(R_2)}.
\]
This monotonicity exposes an important deployment issue, where smaller rerun budgets systematically under-observe flaky failures because some failures that would pass on later reruns remain labeled as persistent under a limited budget.
Hence, the observed label depends not only on the failure behavior itself but also on the rerun policy used to generate supervision.
This one-sided censoring effect motivates the rerun-budget correction proposed later in the paper.

\textit{\textbf{Thresholded decisions}}.
The deployment objective in SCOUT is a thresholded action, not ranking quality alone. Let $\tau \in [0,1]$ be the decision threshold. The policy takes action:
\[
a_i =
\begin{cases}
\texttt{rerun}, & p_i \ge \tau,\\
\texttt{escalate}, & p_i < \tau.
\end{cases}
\]
Suppose that automatically rerunning a persistent failure incurs cost $c_{\mathrm{fp}}$, missing a flaky failure incurs cost $c_{\mathrm{fn}}$, and issuing an automatic rerun itself incurs overhead $c_{\mathrm{auto}}$. Under calibrated probabilities, the Bayes-optimal threshold is
\[
\tau^\ast = \frac{c_{\mathrm{auto}} + c_{\mathrm{fp}}}{c_{\mathrm{fp}} + c_{\mathrm{fn}}}.
\]
\section{Overview  of SCOUT}

\begin{figure}[!t]
\centering
\includegraphics[width=0.48\textwidth]{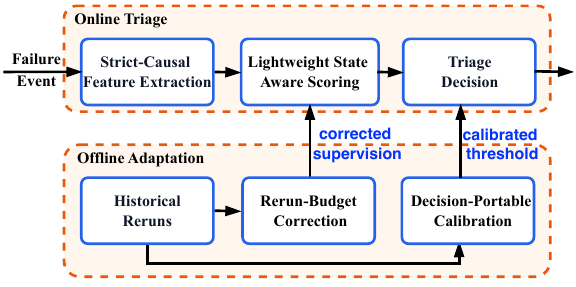}
\caption{The architecture of SCOUT.}
\vspace{-0.3cm}
\label{fig:arch}
\end{figure}

This section gives a high-level introduction to SCOUT.
Figure~\ref{fig:arch} illustrates the overall framework of SCOUT.
At a high level, SCOUT consists of two tightly coupled paths, which are an \emph{online triage} path for immediate rerun-versus-escalate decisions after a failure is observed and an \emph{offline adaptation} path that improves future online decisions using historical rerun results.
The online path contains three components, namely \emph{Strict-Causal Feature Extraction}, \emph{Lightweight State-Aware Scoring}, and \emph{Triage Decision}. The offline path is driven by \emph{historical reruns} and contains two adaptation modules, including \emph{Rerun-Budget Correction} and \emph{Decision-Portable Calibration}.
Together, these components form a deployment-oriented framework in the production environment.

The online path starts from a \emph{Failure Event}, which serves as the trigger signal of SCOUT.
Once a primary CI run fails, the system activates \emph{Strict-Causal Feature Extraction}.
This component is responsible for constructing the model input under a strict-causal protocol.
Concretely, it uses only information available no later than the failure observation time, including pre-failure telemetry, pre-run metadata, and historical statistics computed from earlier primary runs.
By explicitly blocking post-failure artifacts such as error strings, failure logs, and future rerun information, this component ensures that the inference-time feature space is aligned with the intended online deployment setting.

The extracted feature vector is then passed to \emph{Lightweight State-Aware Scoring}, which forms the predictive core of SCOUT.
This component estimates the flaky probability of the failed run from runtime state signals under millisecond-scale, CPU-only serving constraints. 
The design is intentionally lightweight.
SCOUT does not rely on heavy diagnosis pipelines or large multimodal models.
It uses compact models over aggregated telemetry, context, and optional sparse metadata features. 
The goal of this component is not only predictive accuracy but also low-latency, stable, and interpretable online scoring.

The output score is consumed by the \emph{Triage Decision} component. 
Rather than treating the model output as a ranking signal alone, SCOUT maps the score to a concrete operational action using a fixed decision threshold.
In this way, the online path directly implements the deployment objective of the system, which decides whether a failed run should be automatically rerun as likely flaky or escalated as likely persistent.
Therefore, this component operationalizes the probability into an actionable policy.

The offline path continuously improves this online policy using \emph{Historical Reruns}.
This data source records past rerun results and provides the supervision signal for model adaptation.
SCOUT uses it in two ways.
First, the \emph{Rerun-Budget Correction} module addresses the fact that observed flakiness labels are generated under a finite rerun budget.
Since smaller rerun budgets systematically under-observe flaky failures, this module constructs corrected supervision for training, making the scorer more robust to policy-induced label bias.
Its output is fed back to the scoring component as corrected supervision for model updating.

Second, the \emph{Decision-Portable Calibration} module improves the portability of the online decision rule under temporal and cross-domain shifts.
Given historical rerun outcomes, this module calibrates model scores so that a single threshold can remain useful across changing workloads, versions, and environments.
Instead of requiring per-domain retuning, SCOUT uses this module to update the decision threshold in a way that preserves action quality when the score distribution shifts.
Therefore, its output is fed back to the \emph{Triage Decision} component as a calibrated threshold.

Overall, SCOUT separates immediate online inference from slower offline adaptation while connecting them through supervision and decision updates. 
The online path guarantees strict-causal, low-latency triage for each failed run, and the offline path improves both what the system learns from and how it acts on the predicted flaky probability.
This architecture makes SCOUT a practical framework for state-aware flaky-failure triage in a distributed database~CI.
\section{Methodology}

\subsection{Strict-Causal Feature Extraction}\label{subsec:feature_extraction}

\begin{algorithm}[!t]
\caption{Strict-Causal Feature Extraction}
\label{alg:strict_causal_features}
\KwIn{Failed primary run $i$ observed at failure time $t_i$}
\KwOut{Strict-causal feature vector $x_i$}

Initialize $x_i \leftarrow \emptyset$\;

Collect telemetry in the pre-failure window $[t_i-120\text{s},\, t_i]$\;

\ForEach{telemetry family $m \in \mathcal{M}$}{
    collect all observations of $m$ in $[t_i-120\text{s},\, t_i]$\;
    compute window aggregates $\{\mathrm{mean}, \mathrm{max}, \mathrm{std}, \mathrm{p95}\}$\;
    append the aggregates of $m$ to $x_i$\;
}

Extract governed pre-run metadata tokens and append them to $x_i$\;

Compute history features from earlier primary runs only and append them to $x_i$\;

\ForEach{feature $f$ with missing telemetry}{
    impute $f \leftarrow 0$\;
    record the missingness ratio of $f$\;
}

\If{key-feature missingness exceeds the predefined threshold}{
    mark the run as escalation-preferred\;
}

\Return{$x_i$}
\end{algorithm}

The \emph{Strict-Causal Feature Extraction} component of SCOUT constructs the input representation for online triage under a strict-causal deployment protocol.
Let the time of failure be $t=0$.
SCOUT uses only information available no later than this time and blocks all post-failure artifacts from the main pipeline.
As a result, feature extraction is constrained to pre-failure telemetry, pre-run metadata, and historical statistics computed from earlier primary runs.

SCOUT extracts three categories of strict-causal features. The first category is \emph{dense telemetry}, which captures the runtime state immediately before failure. 
The second category is \emph{pre-run metadata tokens}, which provide cheap CI-side context without introducing post-failure leakage. 
The third category is \emph{strict-causal history}, computed only from earlier primary runs of the same test identity. 
This design ensures that the input representation remains causally valid for online rerun-versus-escalate decisions. Algorithm~\ref{alg:strict_causal_features} summarizes the extraction procedure.

For dense telemetry, SCOUT uses a fixed pre-failure window $[-120\text{s},0]$ (line~2).
Within this window, it collects low-level metrics from 14 telemetry families, including lock/transaction, Raft, LSM/storage, resource, and queue signals.
Concretely, these families cover CPU usage, I/O wait, disk latency, network RTT, lock-wait ratio, transaction retry rate, deadlock count, LSM compaction/flush/write-stall, Raft apply/proposal latency and leader changes, and queue wait.
Each metric family is aggregated into window-level summary statistics $\{\mathrm{mean}, \mathrm{max}, \mathrm{std}, \mathrm{p95}\}$ before appending them to the feature vector (lines~3-6).
Compared with raw logs or per-timestep sequences, this aggregation produces a compact state representation that is efficient to compute and stable for lightweight online inference.

SCOUT further augments the dense telemetry representation with strict-causal context and history features.
Pre-run metadata tokens encode cheap CI-side descriptors such as workload, version, and other governed metadata fields (line~7).
Historical features are computed only from earlier primary runs and exclude any lookahead into future reruns or future failures (line~8).
In the current design, these features include the test-duration baseline and the historical failure rate.
Because sparse metadata can easily leak identity, SCOUT treats feature governance as
part of extraction itself. 
Raw identifiers and post-failure artifacts are excluded from the main~protocol.

The component is also designed to be robust to incomplete telemetry.
In telemetry-based deployments, some metrics may be missing or delayed within the extraction window.
Therefore, SCOUT imputes missing aggregates with 0 and records per-feature missing ratios (lines~9-11).
When key features exceed a missingness threshold, the system can disable auto-rerun and fall back to escalation (lines~12-13).
In this way, the component provides a compact, auditable, and deployment-compatible feature vector for the downstream scoring module (line~14).

\subsection{Lightweight State-aware Scoring}

\begin{algorithm}[!t]
\caption{Lightweight State-Aware Scoring}
\label{alg:lightweight_scoring}
\KwIn{Strict-causal feature vector $x_i$, scorer type $s \in \{\textsc{StateLR}, \textsc{TextLR}, \textsc{SDJLR}\}$}
\KwOut{Flaky probability $\hat p_i$}

Split $x_i$ into dense features $x_i^{(d)}$ and sparse metadata tokens $x_i^{(s)}$\;

Standardize dense features to obtain $\tilde{x}_i^{(d)}$\;

Construct the sparse TF-IDF vector $v_i^{(s)}$ from $x_i^{(s)}$\;

\eIf{$s=\textsc{StateLR}$}{
    set $h_i \leftarrow \tilde{x}_i^{(d)}$\;
    compute $\hat p_i \leftarrow \sigma(w^\top h_i)$\;
}{
    \eIf{$s=\textsc{TextLR}$}{
        set $h_i \leftarrow v_i^{(s)}$\;
        compute $\hat p_i \leftarrow \sigma(w^\top h_i)$\;
    }{
        concatenate $h_i \leftarrow [\tilde{x}_i^{(d)}; v_i^{(s)}]$\;
        compute $\hat p_i \leftarrow \sigma(w^\top h_i)$\;
    }
}

\Return{$\hat p_i$}
\end{algorithm}

Lightweight state-aware scoring maps the strict-causal feature vector to a flaky probability for
online triage.
The design goal of this component is to preserve the most actionable pre-failure runtime state while remaining compatible with CPU-only, millisecond-scale deployment.
To this end, SCOUT uses a lightweight logistic scoring family rather than heavy sequence or multimodal models.

SCOUT supports three lightweight scoring variants. The first is a \emph{dense state model}, denoted as \textbf{State LR}, which applies logistic regression to standardized dense telemetry and context features.
The second is a \emph{sparse metadata model}, denoted as \textbf{Text LR}, which operates on a lightweight TF-IDF representation of cheap pre-run CI metadata. 
The third is a \emph{sparse+dense fusion model}, denoted as \textbf{SDJ-LR}, which concatenates the standardized dense features with the sparse metadata representation and fits a single logistic regression model.
Although these variants use different input representations, they share the same scoring form. For a failed run $i$, SCOUT predicts
\[
\hat p_i = \sigma(w^\top h_i),
\]
where $\sigma(\cdot)$ is the logistic function and $h_i$ is the representation used by the selected variant.

Algorithm~\ref{alg:lightweight_scoring} summarizes the scoring procedure.
Given a strict-causal feature vector $x_i$, the scorer first splits it into dense features and sparse metadata tokens (line~1).
It then standardizes the dense channel and constructs the sparse TF-IDF representation for the metadata channel (lines~2-3).
Next, it applies one of three scoring variants: State LR on the dense channel (lines~4-5), Text LR on the sparse channel (lines~6-7), or SDJ-LR on the concatenated sparse+dense representation (lines~8-10). Finally, it returns the predicted flaky probability (line~11).

The dense state model uses standardized telemetry/context features as input and computes
\[
h_i = \tilde{x}_i^{(d)},
\]
where $\tilde{x}_i^{(d)}$ denotes the standardized dense telemetry and context features (lines~4-6).
The purpose of State LR is to capture the transient runtime state directly from the strict-causal dense representation. 
Because the dense features have already been aggregated over the pre-failure window, this model avoids expensive sequence processing while remaining suitable for online inference.

For the sparse metadata variant, SCOUT sets
\[
h_i = v_i^{(s)},
\]
where $v_i^{(s)}$ is the TF-IDF representation of governed pre-run metadata tokens (lines~7-10). 
The sparse channel is not intended to replace telemetry-based state information.
Instead, it provides cheap CI-side context, such as workload, version, and other governed metadata fields, that can complement the dense state representation when available.

For the fusion variant, SCOUT concatenates the two channels (lines~11-13)
\[
h_i = [\tilde{x}_i^{(d)}; v_i^{(s)}].
\]
In the current design, SDJ-LR uses the SAGA logistic predictor over the combined representation, which preserves interpretability and predictable serving cost while allowing dense runtime state and sparse metadata to contribute jointly to the flaky score.

\subsection{Decision-Portable Calibration}

\begin{algorithm}[t]
\caption{Decision-Portable Calibration (OA-Cal)}
\label{alg:decision_portable_calibration}
\KwIn{Base scorer $f$, source calibration set $\mathcal{D}_{\mathrm{src}}^{\mathrm{cal}}$, unlabeled target inputs $\mathcal{X}_{\mathrm{tgt}}$, candidate calibrators $\mathcal{A}$, threshold $\tau^\ast$}
\KwOut{Calibrated mapping $\mathcal{C}^\ast(\cdot)$}

Compute source scores $\{\hat p_j^{\mathrm{src}}\}$ and target scores $\{\hat p_k^{\mathrm{tgt}}\}$ using $f$\;

Fit a domain classifier $g$ on source vs.\ target scores\;

Estimate clipped importance weights and effective sample size (ESS) from $g$\;

\eIf{ESS is low or source-target separability is extreme}{
    disable importance weighting\;
    use uniform weights on the source calibration set\;
}{
    enable importance weighting on the source calibration set\;
}

\ForEach{candidate calibrator $\mathcal{C} \in \mathcal{A}$}{
    perform repeated source-only cross-fits on $\mathcal{D}_{\mathrm{src}}^{\mathrm{cal}}$\;
    fit $\mathcal{C}$ on one split using the selected weights\;
    evaluate fixed-threshold decision cost at $\tau^\ast$ on the held-out split\;
    record the mean cost $\mu_{\mathcal{C}}$ and standard deviation $\sigma_{\mathcal{C}}$\;
}

Select
\[
\mathcal{C}^\ast = \arg\min_{\mathcal{C} \in \mathcal{A}}
\left(\mu_{\mathcal{C}} + \kappa \sigma_{\mathcal{C}}\right)
\]
using Cost-UCB\;

Fit $\mathcal{C}^\ast$ on the full source calibration set with the selected weighting scheme\;

\Return{$\mathcal{C}^\ast(\cdot)$}
\end{algorithm}

The component of SCOUT is to make a single decision threshold transferable across temporal and cross-domain shifts.
As defined in the preliminaries, SCOUT uses a fixed cost-derived threshold $\tau^\ast$ to map a calibrated flaky probability to a rerun-versus-escalate action. 
In SCOUT, the scorer outputs a raw flaky probability $\hat p_i$, but the deployment objective is not ranking quality alone.
Instead, the system must map this score to a concrete rerun-versus-escalate action under a fixed cost model.
Accordingly, SCOUT applies post hoc calibration so that the score used for decision-making is aligned with the deployment threshold.

The main challenge is that a calibrator fitted in one source domain may transfer poorly to a target domain with different workloads, versions, or operating conditions.
In this setting, using target labels to tune calibration is often unrealistic.
Therefore, SCOUT adopts an overlap-aware, source-only calibration procedure, denoted as \textbf{OA-Cal}.
Its design follows three principles.
First, target labels are never used for calibrator selection. 
Second, importance weighting is applied only when source-target overlap is adequate.
Third, calibrator selection is driven by the deployment objective at the fixed threshold $\tau^\ast$, rather than by generic calibration fitness alone.

Algorithm~\ref{alg:decision_portable_calibration} summarizes the procedure. 
Given a trained base scorer, a labeled source calibration set, and unlabeled target inputs, OA-Cal first computes source and target raw scores (line~1).
It then fits a domain classifier on source versus target scores and derives overlap diagnostics, including clipped importance weights and effective sample size (lines~2-3).
When overlap is poor, weighting is disabled, and the procedure falls back to unweighted source calibration (lines~4-8).
Next, OA-Cal evaluates a set of candidate calibrators using repeated source-only cross-fits (lines~9--13).
In each cross-fit, a calibrator is fitted on one split and evaluated on the other split by the fixed-threshold decision cost at $\tau^\ast$, yielding a mean cost and a variance estimate.
Finally, SCOUT selects the calibrator using a Cost-UCB criterion, fits it on the full source calibration set, and returns the resulting calibrated mapping (lines~14-16).

In the current SCOUT design, the candidate set includes lightweight post-hoc mappings such as sigmoid scaling, isotonic regression, beta calibration, and compact tree- or binning-based calibrators.
When the available calibration set is small, or overlap diagnostics are poor, SCOUT favors smoother low-variance mappings and excludes brittle high-variance ones. 
In this way, this component turns the raw scorer output into a deployment-compatible probability that can support portable rerun-versus-escalate decisions without per-target threshold retuning.

\subsection{Rerun-Budget Correction}

\begin{algorithm}[t]
\caption{Rerun-Budget Correction via Posterior-Soft Target}
\label{alg:rerun_budget_correction}
\KwIn{Observed first-$R$ rerun outcomes for run $i$, oracle budget $R' > R$, Beta prior parameters $(\alpha,\beta)$}
\KwOut{Corrected soft target $\tilde y_i^{(R \rightarrow R')}$}

\eIf{at least one pass is observed in the first $R$ reruns}{
    set $\tilde y_i^{(R \rightarrow R')} \leftarrow 1$\;
}{
    update posterior: $q \mid \text{first }R \sim \mathrm{Beta}(\alpha,\beta+R)$\;
    compute
    \[
    m_i \leftarrow \prod_{j=0}^{R'-R-1}
    \frac{\beta+R+j}{\alpha+\beta+R+j}
    \]
    as the posterior probability of no pass in the remaining $R'-R$ reruns\;
    set
    \[
    \tilde y_i^{(R \rightarrow R')} \leftarrow 1 - m_i
    \]
    as the posterior-soft any-pass target\;
}

\Return{$\tilde y_i^{(R \rightarrow R')}$}
\end{algorithm}

The component of SCOUT addresses the label bias induced by finite rerun policies.
As introduced in the preliminaries, the observed flaky label depends on the available rerun budget, i.e., with a smaller budget $R$, some failures that would be labeled flaky under a larger budget $R' > R$ remain labeled as persistent because no pass is observed within the first $R$ reruns. 
Consequently, the training label is policy-generated and one-sidedly censored. 
Therefore, the component replaces the finite-budget hard label with a larger-budget soft target that better reflects the decision-time oracle of interest.

SCOUT adopts a simple posterior-imputation strategy.
Let $y_R$ denote the any-pass label under rerun budget $R$, and let $R' > R$ denote the oracle rerun budget used as the correction target. 
Suppose the first $R$ reruns are observed. If at least one pass already appears in this prefix, then the larger-budget label is deterministically positive, and the soft target is set to 1.
Otherwise, all observed reruns fail, and SCOUT places a Beta prior on the per-rerun pass probability $q \sim \mathrm{Beta}(\alpha,\beta)$. After observing an all-fail prefix of length $R$, the posterior becomes
\[
q \mid \text{first }R \sim \mathrm{Beta}(\alpha,\beta+R).
\]
The posterior-soft target for the larger-budget any-pass label is then
\[
\tilde y_i^{(R \rightarrow R')}
= P(y_{R'}=1 \mid \text{first }R)
= 1 - E[(1-q)^{R'-R}],
\]
which has the closed form
\[
\tilde y_i^{(R \rightarrow R')}
= 1-\prod_{j=0}^{R'-R-1}
\frac{\beta+R+j}{\alpha+\beta+R+j}
\]
in the all-fail case. Thus, SCOUT replaces the censored hard label with the posterior probability that a larger rerun budget would observe at least one pass.

Algorithm~\ref{alg:rerun_budget_correction} summarizes the correction procedure. 
Given the first $R$ rerun outcomes and an oracle budget $R'$, SCOUT first checks whether any pass is already observed in the prefix (lines~1-2).
If so, the posterior-soft target is set to 1.
Otherwise, it updates the Beta posterior under the all-fail prefix, computes the probability that no pass occurs in the remaining $R'-R$ reruns via the closed-form Beta moment (lines~3-5), and converts it to the posterior-soft any-pass target (line~6).
Finally, the resulting corrected supervision is returned for model training (line~7).
This is why the correction is both explicit and lightweight. It reduces to a small number of scalar operations per example.

In the current design, the prior parameters $(\alpha,\beta)$ are estimated by empirical Bayes on the training set using a method-of-moments beta-binomial fit.
This keeps the correction tied to the observed rerun regime while avoiding example-specific latent variable inference.
Importantly, the corrected target is used only in the offline adaptation path.
It does not alter the online decision protocol, but instead changes the supervision used to update the lightweight scorer.
\section{Experiment Evaluation}

To evaluate the effectiveness of SCOUT, we aim to answer the following questions:

\begin{itemize}
    \item \textbf{RQ1:} How effective is SCOUT for flaky-failure triage compared with lightweight history-, metadata-, tree-, and sequence-based baselines on strict causal features?
    \item \textbf{RQ2:} Is calibration, rather than heavier model classes, the key to portable fixed-threshold decisions under temporal and cross-domain shift?
    \item \textbf{RQ3:} Does rerun-budget correction mitigate the label bias induced by finite rerun policies and improve learning toward a larger-budget oracle label?
    \item \textbf{RQ4:} Is SCOUT practical for deployment under strict-causal, CPU-only online triage constraints?
    \item \textbf{RQ5:} Do the main conclusions of SCOUT transfer to real services and real CI traces beyond the synthetic benchmark?
\end{itemize}

\subsection{Experimental Setup}

\subsubsection{Benchmarks and Data Sources}

Our main benchmark contains 3,680 labeled failed runs, including 462 flaky positives (12.55\%), and 62 telemetry/context features.
It is generated by a controllable simulator of the distributed database CI with 12,000 primary runs. For each primary run, we sample a workload type, test identity, database version, and fault condition. The simulator covers representative fault families, including network delay, packet loss, CPU pressure, disk I/O noise, and leader change, together with a no-fault case. Each fault is assigned a severity level and translated into structured perturbations over pre-failure telemetry, such as network RTT, retry rate, CPU utilization, disk latency, and Raft-related indicators. These perturbations affect both the primary-run failure probability and a run-specific rerun-pass probability. For every fail-like primary run, we simulate a rerun budget of $R=8$ and assign a flaky label if at least one rerun succeeds; otherwise, the failure is labeled persistent. We use this benchmark as the primary testbed for studying strict causal feature extraction, distribution shift, fixed-threshold decision portability, and rerun-policy-induced label bias.

To reduce over-alignment to a fixed simulator feature table, we also construct a fully generative synthetic benchmark in which dense telemetry is sampled from a nonlinear manifold with localized failure modes and temporal drift in the latent mixture. This benchmark is used as a stress test for whether the main conclusions continue to hold under a more flexible data-generating process.

For real-service feasibility evaluation, we deploy TiDB v7 and v8 clusters in Docker together with Prometheus monitoring and execute controlled SQL workloads under both container-level and workload-level fault injection.
The container-level faults include stopping the TiDB service and restarting a TiKV node, while the workload-level failures include timeout-inducing sleeps, write conflicts, lock-wait timeouts, and deadlocks. For each failed primary run, we perform a rerun budget of  $R=5$ and assign a rerun-based flaky label using the same any-pass rule as in the main benchmark.
This produces a real-service dataset of 1,600 runs, 341 labeled failures, and 194 positives under the rerun-based label.
Telemetry is collected from both Prometheus queries and Docker container statistics.

For large-scale CI validation, we collect public GitHub Actions workflow-run metadata, which has 157,807 workflow runs (157,639 completed) from 36 repositories spanning distributed databases and data systems.
We derive SHA-group failure episodes by grouping runs by \texttt{(repo, workflow\_id, head\_sha)}, taking the first completed non-success run as the primary failure, and searching for up to three subsequent completed runs within a 24-hour horizon.
This yields 36,241 episodes, including 3,953 labeled-valid episodes with at least one observed rerun and 32,288 censored episodes with no observed rerun.
An episode is labeled valid if at least one subsequent run is observed; otherwise, it is treated as censored.
For labeled-valid episodes, we assign a rerun-based label indicating whether any observed subsequent run succeeds.
Features in this benchmark are strictly metadata-only, including workflow/event/branch/timestamp information and strict causal history features.

\subsubsection{Environment}
All experiments are conducted in a CPU-only setting, consistent with the deployment target of SCOUT.
For latency evaluation, we use an Apple M4 machine as the main local measurement platform and additionally report a conservative \texttt{Linux/AMD64} environment.
This separation allows us to distinguish development-time latency from a more deployment-like setting. 
The online path is evaluated under repeated calls so that the reported latency reflects steady-state serving behavior rather than one-time initialization cost.

\subsection{Effectiveness on Strict-Causal Features (RQ1)}

\begin{table}[!t]
\centering
\caption{Temporal split on strict-causal features (window $[-120\text{s},0]$). Sparse baselines use metadata-only tokens (no injected fault tags). Seq LR flattens per-metric per-timestep telemetry ($5\times14=70$ dims). Rows shaded in light orange and marked with $\dagger$ denote SCOUT scoring variants; unshaded rows are baselines.}
\label{tab:rq1_temporal_main}
\small
\setlength{\tabcolsep}{6pt}
\begin{tabular}{lcccc}
\toprule
Model & PR-AUC & ROC-AUC & ECE(10) & Brier \\
\midrule
History+churn LR                 & 0.124 & 0.535 & 0.386 & 0.247 \\
\rowcolor{scoutscore}
State LR (dense)$^\dagger$       & 0.199 & 0.637 & 0.345 & 0.226 \\
Seq LR (timepoint, dense)        & 0.179 & 0.639 & 0.346 & 0.226 \\
\rowcolor{scoutscore}
Text LR (TF-IDF)$^\dagger$       & 0.123 & 0.515 & 0.363 & 0.245 \\
\rowcolor{scoutscore}
SDJ-LR (dense+sparse)$^\dagger$  & 0.206 & 0.628 & 0.324 & 0.221 \\
HistGB (tuned sweep)             & \textbf{0.213} & \textbf{0.645} & 0.339 & 0.214 \\
RF (tuned sweep)                 & 0.194 & 0.613 & \textbf{0.207} & \textbf{0.149} \\
LightGBM (tuned sweep)           & 0.199 & 0.627 & 0.261 & 0.180 \\
\bottomrule
\end{tabular}
\end{table}

\begin{table}[!t]
\centering
\caption{Temporal 60/20/20: fixed-$\tau^\ast$ decision cost across models after post-hoc calibration (lower is better). SDJ-LR uses governed sparse metadata tokens (no commit/test identifiers). Rows shaded in light orange and marked with $\dagger$ denote SCOUT scoring variants; unshaded rows are baselines.}
\label{tab:rq1_cost_family}
\small
\setlength{\tabcolsep}{1pt}
\begin{tabular}{lcccc}
\toprule
Model & PR-AUC & Uncal cost & Sigmoid cost & Isotonic cost \\
\midrule
\rowcolor{scoutscore}
State LR (dense)$^\dagger$                 & 0.191 & 765.4 & 538.4 & \textbf{496.9} \\
Spline GAM (dense)                         & 0.236 & 766.4 & 531.0 & 503.0 \\
Tiny GRU (dense seq)                       & 0.197 & 765.4 & 544.5 & 521.2 \\
Tiny CNN (dense seq)                       & 0.201 & 765.4 & \textbf{523.0} & 558.9 \\
\rowcolor{scoutscore}
SDJ-LR (dense+sparse, no-id)$^\dagger$     & 0.196 & 764.2 & 536.0 & 508.0 \\
HistGB (dense)                             & 0.209 & 765.4 & 546.1 & 557.2 \\
RF (dense)                                 & 0.184 & \textbf{756.9} & 570.0 & 547.0 \\
LightGBM (dense)                           & 0.215 & 766.9 & 575.5 & 534.4 \\
\bottomrule
\end{tabular}
\end{table}

\subsubsection{Setup}
We evaluate the main triage task on the synthetic benchmark of 3,680 labeled failed runs.
All models use only pre-failure telemetry from the window $[-120\text{s},0]$, together with governed pre-run metadata and history features computed only from earlier primary runs.
For the main ranking comparison, we use a temporal split that trains on the earliest 80\% of the runs and tests on the latest 20\%.
We report PR-AUC as the primary metric under class imbalance, together with ROC-AUC, ECE(10), and Brier score.
We compare SCOUT against several lightweight baselines and model families.
These include a \emph{History+churn LR} baseline with four cheap features, \emph{State LR} on dense strict-causal telemetry/context features, \emph{Seq LR} on flattened per-timestep telemetry, \emph{Text LR} on sparse metadata-only tokens, and \emph{SDJ-LR} as sparse+dense fusion.
We further include tuned tree baselines, namely HistGB, Random Forest (RF), and LightGBM, to test whether slightly heavier tabular models provide substantial additional value under the same strict causal feature extraction.

\subsubsection{Results}

Table~\ref{tab:rq1_temporal_main} reports the temporal-split ranking results under the strict-causal window.
The \emph{History+churn LR} is weak (PR-AUC 0.124), and the metadata-only \emph{Text LR} is similarly limited (PR-AUC 0.123), indicating that cheap history-only or metadata-only signals are insufficient in this setting.
In contrast, dense strict-causal telemetry substantially improves triage quality, i.e., \emph{State LR} reaches PR-AUC 0.199 and ROC-AUC 0.637.
The sequence-aware but still \emph{lightweight Seq LR} baseline underperforms the window-aggregated dense model (PR-AUC 0.179 vs.\ 0.199), suggesting that simple pre-failure window statistics capture most of the usable dynamics for online triage.
Adding governed sparse metadata on top of dense telemetry yields a modest improvement, i.e., \emph{SDJ-LR} reaches PR-AUC 0.206 and improves over both dense-only and sparse-only logistic baselines.

Among the tuned tree models, \emph{HistGB} achieves the highest PR-AUC (0.213), followed by \emph{LightGBM} (0.199) and RF (0.194).
However, the gain over SDJ-LR is small in absolute terms, and the tree models do not provide a decisive advantage in the ranking setting.
This already suggests that low-level runtime telemetry is the main source of useful features, while model-class complexity contributes only marginal additional benefit.

To evaluate whether this observation still holds when decisions are made through a fixed global threshold, we further compare model families under a disjoint 60/20/20 train/calibration/test split. 
As shown in Table~\ref{tab:rq1_cost_family}, uncalibrated costs are uniformly high across model classes.
In particular, isotonic-calibrated \emph{State LR} attains the lowest decision cost (496.9), outperforming calibrated tree baselines such as \emph{HistGB} (557.2), \emph{RF} (547.0), and \emph{LightGBM} (534.4).
\emph{SDJ-LR} remains competitive after calibration (508.0), while the ms-feasible temporal encoders \emph{Tiny GRU} and \emph{Tiny CNN} do not improve fixed-threshold decision cost despite slightly stronger sequence-modeling capacity.

Under the strict-causal protocol, SCOUT achieves effective flaky-failure triage with lightweight state-aware models.
Dense pre-failure telemetry provides the dominant source of actionable features, clearly outperforming history-only and metadata-only alternatives, while simple window-level aggregation already captures most of the useful runtime dynamics.
Sparse metadata contributes only a modest complementary gain when fused with dense telemetry. Although slightly heavier models may offer small ranking improvements, these gains do not translate into clear advantages in deployment-oriented fixed-threshold decisions.
Overall, the results validate the core design choice of SCOUT, i.e.,  practical online triage is better served by strict causal runtime state and lightweight scoring than by increasing model complexity.

\subsection{Decision Portability Under Shift (RQ2)}

\begin{table}[t]
\centering
\caption{Calibration-transfer baselines for decision portability at a single global threshold $\tau^\ast$ (lower is better). Cells report cost/ECE(10) on the target domain at fixed $\tau^\ast$. OACal uses overlap-gated weighting and source-only Cost-UCB selection. Rows shaded in light green and marked with $\dagger$ denote the SCOUT-specific calibration method; unshaded rows are baselines.}
\label{tab:rq2_calibration_transfer}
\small
\setlength{\tabcolsep}{1pt}
\begin{tabular}{lcccc}
\toprule
Method & Temporal (60/20/20) & Cross-workload & $v7.x \rightarrow v8$ & $v8 \rightarrow v7.5$ \\
\midrule
Uncal        & 765.4/0.337 & 1224.5/0.336 & 1244.9/0.326 & 1290.5/0.323 \\
Sigmoid      & 538.4/0.006 & 869.5/0.025  & 1153.9/0.041 & 930.6/0.007  \\
BetaCal      & 524.7/0.016 & \textbf{839.3}/\textbf{0.024} & 995.9/0.038  & \textbf{930.5}/0.023 \\
Isotonic     & \textbf{496.9}/0.017 & 870.5/0.035  & 875.5/0.033  & 998.8/0.066  \\
CalibTree    & 532.5/0.025 & 1004.5/0.062 & 934.9/0.045  & 1091.4/0.073 \\
Venn--Abers  & 501.5/0.019 & 867.4/0.029  & \textbf{867.9}/\textbf{0.018} & 1012.8/0.056 \\
BBQ-lite     & 556.7/0.034 & 941.0/0.033  & 899.9/0.027  & 985.9/0.034  \\
Cluster-TS   & 748.9/0.310 & 1224.5/0.317 & 1246.0/0.326 & 1280.2/0.309 \\
\rowcolor{scoutadapt}
OA-Cal$^\dagger$ & \textbf{496.9}/0.017 & \textbf{839.3}/\textbf{0.024} & 875.5/0.033 & \textbf{930.5}/0.023 \\
\bottomrule
\end{tabular}
\end{table}

\begin{table}[t]
\centering
\caption{OA-Cal gate ablations for decision portability (cells show cost/ECE(10) at fixed $\tau^\ast$). The light-green column marked with $\dagger$ denotes the final SCOUT calibration method; unshaded columns are ablations of OA-Cal.}
\label{tab:rq2_oacal_ablation}
\label{tab:rq2_cost_robust}
\small
\setlength{\tabcolsep}{2pt}
\begin{tabular}{lcccc}
\toprule
Scenario & $n_{\mathrm{cal}}$ & \cellcolor{scoutadapt}OA-Cal$^\dagger$ & +force weights & +allow isotonic \\
\midrule
Temporal         & 736 & \cellcolor{scoutadapt}496.9/0.017 & 496.9/0.017   & 496.9/0.017 \\
Cross-workload   & 497 & \cellcolor{scoutadapt}\textbf{839.3}/\textbf{0.024} & 1004.5/0.062 & 870.5/0.035 \\
$v8 \rightarrow v7.5$ & 246 & \cellcolor{scoutadapt}930.5/0.023 & 930.5/0.023 & 930.5/0.023 \\
\bottomrule
\end{tabular}
\end{table}

\subsubsection{Setup}
To isolate the effect of calibration from the effect of model-family choice, we fix the base scorer to the dense \emph{State LR} model, which attains the lowest calibrated fixed-threshold decision cost among the model families evaluated in Table~\ref{tab:rq1_cost_family}.
We compare the following calibration baselines: \emph{Uncal}, \emph{Sigmoid}, \emph{BetaCal}, \emph{Isotonic}, \emph{CalibTree}, \emph{Venn--Abers}, \emph{BBQ-lite}, \emph{Cluster-TS}, and the proposed \emph{OA-Cal}.
Evaluation is performed in four settings: a temporal 60/20/20 split, a cross-workload transfer, and two cross-version transfers ($v7.x \rightarrow v8$ and $v8 \rightarrow v7.5$). 
In all cases, a single global threshold $\tau^\ast$ is fixed from the deployment cost model and is not re-tuned on the target domain.
We report fixed-threshold decision cost at $\tau^\ast$ and ECE(10) as the primary metrics. 
For OA-Cal, target labels are never used. The calibrator selection is based only on the source calibration split and unlabeled target scores, using overlap diagnostics, optional importance weighting, and Cost-UCB selection.

\subsubsection{Results}

Table~\ref{tab:rq2_calibration_transfer} shows that calibration is already essential in the in-domain temporal setting.
On the temporal 60/20/20 split, the uncalibrated scorer incurs a decision cost of 765.4 with ECE 0.337.
All calibrators reduce this cost substantially, with isotonic reaching 496.9/0.017 and OA-Cal matching the same result by selecting isotonic on the large calibration set.
Thus, even before cross-domain transfer is considered, raw scores are not reliable enough to support a fixed-threshold deployment rule.

The effect becomes stronger under a shift.
In the cross-workload setting, the uncalibrated decision cost rises to 1224.5 with ECE 0.336, showing that naive threshold reuse is unsafe.
The best portable cost is 839.3/0.024, achieved by BetaCal and matched by OA-Cal, while isotonic degrades to 870.5/0.035 and CalibTree further degrades to 1004.5/0.062.
Cluster-TS fails to improve over the uncalibrated score at all.
In the $v7.x \rightarrow v8$ transfer, Venn-Abers achieves the lowest cost (867.9/0.018), while OA-Cal matches isotonic at 875.5/0.033. 
In the harder $v8 \rightarrow v7.5$ transfer, smoother mappings are clearly preferable.
OA-Cal attains 930.5/0.023, matching the best portable cost and improving substantially over isotonic (998.8/0.066). 
Taken together, these results show that no single calibrator dominates across all shifts, but well-chosen calibration is consistently more important than naive threshold reuse.

\begin{figure}[!t]
\centering
\includegraphics[width=0.92\linewidth]{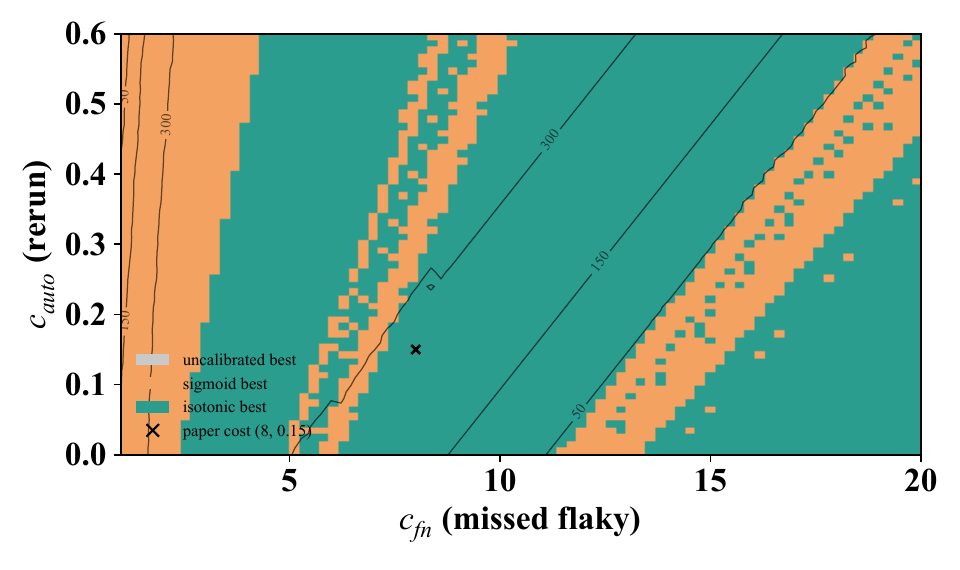}
\Description{Heatmap over cost parameters showing which calibration method yields the lowest fixed-threshold decision cost; isotonic dominates most regions and uncalibrated scores are never optimal.}
\caption{Cost dominance map over \((c_{\text{fn}},c_{\text{auto}})\) with \(c_{\text{fp}}{=}1\) on the temporal 60/20/20 split. Uncalibrated scores are never best; isotonic dominates in 71\% of grid points and sigmoid in 29\%.}
\label{fig:cost-dominance}
\label{fig:rq2_oacal_sensitivity}
\end{figure}
\begin{table}[t]
\centering
\caption{Robustness to cost misspecification at fixed \(\tau^*\) on the temporal 60/20/20 split (\(c_{\text{fp}}{=}1\)). Even under 2\(\times\) errors in \(c_{\text{fn}}\) or \(c_{\text{auto}}\), calibrated probabilities typically reduce decision cost relative to uncalibrated scores.}
\label{tab:cost-misspec}
\small
\begin{tabular}{lrrrr}\toprule
Scenario & $\tau^*$ & Uncal & Sigmoid & Isotonic \\
\midrule
baseline & 0.128 & 765.4 & 538.4 & 496.9 \\
$c_{fn}/2$ & 0.230 & 698.4 & 324.0 & 305.9 \\
$2\times c_{fn}$ & 0.068 & 765.4 & 770.9 & 757.1 \\
$2\times c_{auto}$ & 0.144 & 874.5 & 533.5 & 537.1 \\
\bottomrule\end{tabular}

\end{table}

\begin{figure*}[!t]
\centering
\includegraphics[width=0.98\textwidth]{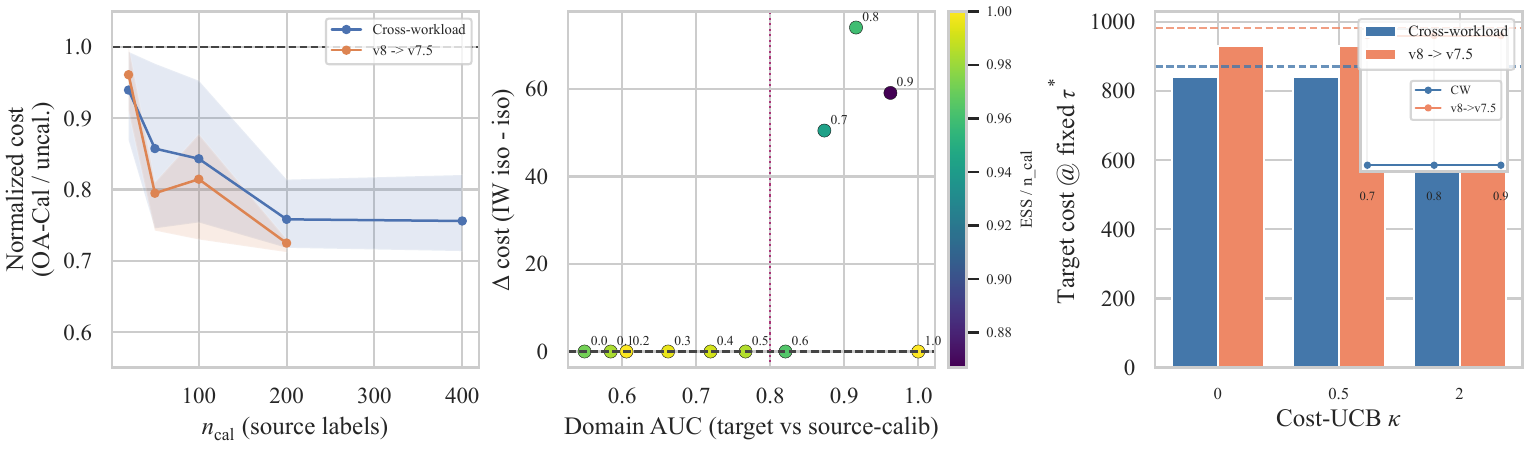}
\Description{Three-panel sensitivity analysis for calibration under shift: left shows decision-cost versus calibration-set size, middle varies source-target overlap and reports effective sample size, and right varies a hyperparameter for a cost-aware calibrator selector.}
\caption{OA-Cal sensitivity for decision portability. Left: label-scarce learning curve (OA-Cal normalized fixed-\(\tau^*\) cost versus calibration-set size \(n_{\text{cal}}\)) for two transfer scenarios. Middle: controlled overlap sweep (mix target with source; annotate mix fraction; color = ESS/\(n_{\text{cal}}\)); importance weighting helps only under moderate overlap. Right: fixed-\(\tau^*\) cost versus Cost-UCB \(\kappa\) (bars) with an inset sweeping the AUC gate threshold.}
\label{fig:oacal-sens}
\label{fig:rq2_cost_map}
\end{figure*}

In the cross-workload setting, forcing importance weighting under extreme source-target separability sharply worsens portability, raising cost/ECE from 839.3/0.024 to 1004.5/0.062 as shown in Table~\ref{tab:rq2_oacal_ablation}.
Allowing isotonic in the same poor-overlap setting also increases cost from 839.3 to 870.5.
These ablations show that it is necessary to avoid weighting and high-variance calibrators when transfer is difficult.
By contrast, in the temporal split and the moderate-overlap $v8 \rightarrow v7.5$ setting, the OA-Cal gates do not hurt because the procedure naturally falls back to the safer choice.
Figure~\ref{fig:rq2_oacal_sensitivity} further confirms this pattern. The importance weighting helps only under moderate overlap and becomes harmful under extreme separability, while the cost-UCB parameter is stable over a broad range.

The calibration benefit is also robust to moderate misspecification of the cost model.
Figure~\ref{fig:rq2_cost_map} shows that uncalibrated scores are never optimal over the explored $(c_{\mathrm{fn}}, c_{\mathrm{auto}})$ grid, whereas isotonic dominates most of the grid and sigmoid dominates the remainder.
Table~\ref{tab:rq2_cost_robust} shows the same trend numerically.
On the temporal split, calibrated scores remain preferable under 2$\times$ perturbations of $c_{\mathrm{fn}}$ or $c_{\mathrm{auto}}$.
For example, when $c_{\mathrm{fn}}/2$, isotonic reduces decision cost from 698.4 to 305.9; when $2 \times c_{\mathrm{auto}}$, it reduces cost from 874.5 to 537.1.
These results indicate that the gain from calibration is not an artifact of a single threshold choice.

Calibration is the key to portable fixed-threshold decisions under shift.
The main performance gains come from transforming raw scores into reliable probabilities, not from further increasing model complexity.
However, no single calibrator is uniformly best across all domains.
The isotonic is strongest with a large in-domain calibration set, whereas smoother mappings are safer under scarce calibration data or poor source-target overlap.
OA-Cal captures this trade-off without using target labels by gating importance weighting and selecting low-variance calibrators when transfer is risky.
Therefore, SCOUT should treat calibration as part of the deployment policy itself, rather than as an optional post-processing step.

\subsection{Rerun-budget Correction (RQ3)}\

\begin{table}[t]
\centering
\caption{Selective-labeling stress test (synthetic): only some error signatures receive reruns/labels. We compare naive training on the censored $y_3$ label against posterior-soft correction with an unweighted empirical-Bayes (EB) prior and an IPW-corrected EB prior. Lower ECE is better; evaluation is against the oracle $y_8$ label. Columns shaded in light green and marked with $\dagger$ denote SCOUT-specific rerun-budget correction variants.}
\label{tab:rq3_selective_labeling}
\small
\setlength{\tabcolsep}{1pt}
\begin{tabular}{lccc}
\toprule
Selection & Naive ECE & \cellcolor{scoutadapt}Posterior-soft ECE$^\dagger$ & \cellcolor{scoutadapt}Posterior-soft+IPW ECE$^\dagger$ \\
\midrule
none   & 0.338 & \cellcolor{scoutadapt}0.010 & \cellcolor{scoutadapt}0.010 \\
mild   & 0.304 & \cellcolor{scoutadapt}0.010 & \cellcolor{scoutadapt}0.014 \\
strong & 0.377 & \cellcolor{scoutadapt}0.022 & \cellcolor{scoutadapt}0.031 \\
\bottomrule
\end{tabular}
\end{table}

\begin{table}[t]
\centering
\caption{Rerun-budget correction on the synthetic benchmark: simulate $R=3$ labels and evaluate against the oracle $R=8$ any-pass label. The empirical-Bayes prior is $\alpha=0.79$ and $\beta=42.20$. Cost is measured at the fixed $\tau^\ast$ induced by $(c_{\mathrm{fp}}, c_{\mathrm{fn}}, c_{\mathrm{auto}})=(1,8,0.15)$. Rows shaded in light green and marked with $\dagger$ denote the SCOUT rerun-budget correction method; unshaded rows are baselines.}
\label{tab:rq3_main_correction}
\small
\setlength{\tabcolsep}{1pt}
\begin{tabular}{lcccc}
\toprule
Method & ECE(10) & Brier & Cost@$\,\tau^\ast$ & Auto@$\,\tau^\ast$ \\
\midrule
Naive train on $y_3$                & 0.320 & 0.213 & 757.4 & 0.99 \\
\rowcolor{scoutadapt}
Posterior-soft target (ours)$^\dagger$ & \textbf{0.027} & 0.096 & 591.7 & 0.49 \\
Direct-$q$ GLM (binom)              & 0.028 & 0.096 & 525.5 & 0.38 \\
Direct-$q$ (early-stop naive)       & 0.028 & \textbf{0.095} & \textbf{521.0} & 0.37 \\
Discrete hazard (censored@$R=3$)    & 0.039 & 0.096 & 551.5 & 0.44 \\
PU correction (Elkan--Noto)         & 0.598 & 0.495 & 765.4 & 1.00 \\
\bottomrule
\end{tabular}
\end{table}

\begin{table}[t]
\centering
\caption{Rerun-budget correction stress tests (mean over 5 seeds): sticky Markov correlation ($\rho$), attempt drift (cooldown/backoff vs.\ contention), an intervention regime, and a hidden-state warm-cache process. Columns shaded in light green and marked with $\dagger$ denote SCOUT posterior-soft correction results.}
\label{tab:rq3_stress_tests}
\scalebox{0.9}{
\small
\setlength{\tabcolsep}{0.2pt}
\begin{tabular}{lccc}
\toprule
Scenario & Naive ECE & \cellcolor{scoutadapt}Posterior-soft ECE$^\dagger$ & \cellcolor{scoutadapt}Posterior-soft Brier$^\dagger$ \\
\midrule
$\rho = 0.0$                     & 0.309 & \cellcolor{scoutadapt}0.017 & \cellcolor{scoutadapt}0.105 \\
$\rho = 0.3$                     & 0.329 & \cellcolor{scoutadapt}0.019 & \cellcolor{scoutadapt}0.085 \\
$\rho = 0.6$                     & 0.355 & \cellcolor{scoutadapt}0.008 & \cellcolor{scoutadapt}0.059 \\
cooldown ($\gamma=0.01$)         & 0.139 & \cellcolor{scoutadapt}0.045 & \cellcolor{scoutadapt}0.225 \\
contention ($\gamma=0.01$)       & 0.309 & \cellcolor{scoutadapt}0.023 & \cellcolor{scoutadapt}0.037 \\
intervention ($p=0.3,\Delta=+0.2$) & 0.161 & \cellcolor{scoutadapt}0.026 & \cellcolor{scoutadapt}0.219 \\
warm-cache ($p=0.4,\Delta=+0.2$) & 0.243 & \cellcolor{scoutadapt}0.069 & \cellcolor{scoutadapt}0.197 \\
\bottomrule
\end{tabular}
}
\end{table}

\begin{table}[t]
\centering
\caption{Real-service rerun censoring on TiDB: simulate $R=3$ labels and evaluate decisions against the oracle $R=5$ any-pass label. Rows shaded in light green and marked with $\dagger$ denote the SCOUT rerun-budget correction method; unshaded rows are baselines.}
\label{tab:rq3_docker_correction}
\small
\setlength{\tabcolsep}{1pt}
\begin{tabular}{lccc}
\toprule
Method & Cost@$\,\tau^\ast$ & ECE(10) & Brier \\
\midrule
Naive train on $y_3$                  & 60.0 & 0.195 & 0.269 \\
\rowcolor{scoutadapt}
Posterior-soft target$^\dagger$       & \textbf{45.8} & 0.157 & 0.257 \\
Discrete hazard (censored@$R=3$)      & 52.0 & \textbf{0.106} & \textbf{0.233} \\
\bottomrule
\end{tabular}
\end{table}

\subsubsection{Setup}
We simulate finite-budget supervision by truncating rerun labels to $R=3$ and evaluate all methods against the larger-budget oracle label $y_8$ on the benchmark.
To isolate the effect of label correction, we keep the downstream strict-causal scoring pipeline fixed and vary only how the training target is constructed. 
We compare the baselines, including naive training on $y_3$, the proposed posterior-soft target, a direct-$q$ Binomial GLM, a direct-$q$ early-stop naive variant, a discrete hazard model for censored reruns, and a PU correction baseline.
We report ECE(10), Brier score, and fixed-threshold decision cost at $\tau^\ast$, together with the auto-rerun rate when relevant.
The empirical-Bayes prior used by the main posterior-soft model is $\alpha=0.79$ and $\beta=42.20$.

To stress-test robustness beyond the main synthetic setting, we further evaluate three additional scenarios.
First, we simulate \emph{selective labeling}, where reruns are observed only for some error signatures, and compare posterior-soft correction with and without IPW-corrected empirical Bayesian priors.
Second, we evaluate \emph{correlated rerun} and \emph{attempt-drift} regimes, including sticky Markov dependence, cooldown/contention drift, intervention, and warm-cache hidden-state processes. 
Third, we test \emph{real-service rerun censoring} on the TiDB trace by simulating $R=3$ labels and evaluating against the oracle $R=5$ any-pass label.

\subsubsection{Results}

Table~\ref{tab:rq3_selective_labeling} shows that posterior-soft correction remains highly stable under selective labeling. 
Across none, mild, and strong selection regimes, naive training on the censored $y_3$ label yields ECE values of 0.338, 0.304, and 0.377, whereas posterior-soft reduces these to 0.010, 0.010, and 0.022, respectively.
The IPW-corrected empirical-Bayes variant is also effective but does not improve over the unweighted posterior-soft baseline and becomes slightly worse under stronger selection. 
This suggests that the main benefit comes from correcting the finite-budget label semantics itself, rather than from additional reweighting.

Table~\ref{tab:rq3_main_correction} reports the main synthetic comparison against the oracle $y_8$ label.
The proposed posterior-soft target substantially improves calibration relative to naive training on $y_3$, reducing ECE from 0.320 to 0.027 and Brier score from 0.213 to 0.096, while also lowering fixed-threshold decision cost from 757.4 to 591.7.
Direct-$q$ alternatives achieve slightly lower decision cost (525.5 and 521.0), but only when the first-$R$ pass-count information is fully observed and the Binomial count model is appropriate. 
In contrast, many CI systems stop on the first pass, which censors rerun counts and makes direct-$q$ methods less natural without additional attempt-level instrumentation or censored-likelihood modeling. 
The discrete hazard baseline is also competitive, but the PU correction baseline performs very poorly, indicating that the finite-rerun problem is not well captured by a generic positive-unlabeled treatment.

The robustness results in Table~\ref{tab:rq3_stress_tests} further support the correction design.
Across sticky Markov dependence ($\rho \in \{0.0, 0.3, 0.6\}$), attempt-drift regimes, intervention, and warm-cache hidden-state processes, posterior-soft consistently yields much lower ECE than the naive censored-label baseline.
In particular, under stronger correlation ($\rho=0.6$), naive ECE rises to 0.355, whereas posterior-soft drops to 0.008.
This indicates that the method remains effective even when reruns are not i.i.d., which matches the intended use of posterior-soft as a lightweight first-order correction rather than a brittle exact model.

Finally, Table~\ref{tab:rq3_docker_correction} shows that the same pattern extends to a real-service trace.
On TiDB, posterior-soft reduces fixed-threshold decision cost from 60.0 to 45.8 and lowers ECE from 0.195 to 0.157 relative to naive training on $y_3$.
The hazard baseline attains the lowest ECE and Brier score on this trace, but its decision cost remains higher than posterior-soft (52.0 vs.\ 45.8).
Thus, posterior-soft provides the best fixed-threshold action quality on the real-service correction task, even when a more specialized censoring model can achieve a slightly better calibration fit.

Rerun-budget correction mitigates the label bias induced by finite rerun policies, and the proposed posterior-soft target is an effective default correction for SCOUT.
It markedly improves calibration relative to naive finite-budget labels and remains robust under selective labeling, correlated reruns, and a real-service trace.
Although direct-$q$ and hazard-based alternatives can be competitive or even slightly better on some metrics when stronger assumptions or richer attempt-level observations are available, posterior-soft is the most broadly compatible correction for SCOUT because it aligns with any-pass label semantics and remains well-defined under early-stop censoring.
These results justify using posterior-soft correction as the default supervision update mechanism in the offline adaptation path.

\subsection{Deployment Practicality (RQ4)}

\subsubsection{Setup}
We evaluate deployment practicality from three perspectives.
First, we evaluate leakage robustness by enforcing disjoint splits and removing identifier tokens from the sparse channel, and by additionally opening a post-failure window only in an explicit leakage ablation.
Second, we measure latency in both a local Apple M4 and a conservative \texttt{Linux/AMD64} setting, reporting end-to-end online latency as well as model-only inference latency.
Third, we assess interpretability using global feature-attribution summaries, coefficient-sign inspection, and a face-validity perturbation check on canonical state features.

\subsubsection{Results}

\begin{table}[!t]
\centering
\caption{Leakage guard evaluation: PR-AUC under disjoint splits with identifier tokens removed. Rows shaded in light orange and marked with $\dagger$ denote SCOUT scoring variants.}
\label{tab:rq4_leakage_guard}
\small
\setlength{\tabcolsep}{8pt}
\begin{tabular}{lccc}
\toprule
Split & \cellcolor{scoutscore}State LR$^\dagger$ & \cellcolor{scoutscore}Text LR$^\dagger$ & \cellcolor{scoutscore}SDJ-LR$^\dagger$ \\
\midrule
Temporal 80/20 & 0.199 & 0.107 & 0.197 \\
Leave-one-\texttt{test\_id}-out & 0.241 & 0.114 & 0.237 \\
Leave-one-\texttt{commit\_id}-out & 0.239 & 0.128 & 0.239 \\
\bottomrule
\end{tabular}
\end{table}

Table~\ref{tab:rq4_leakage_guard} reports the leakage-guard evaluation.
Under the strict temporal 80/20 split, the SCOUT scoring variants achieve PR-AUCs of 0.199 (State LR), 0.107 (Text LR), and 0.197 (SDJ-LR).
Under leave-one-\texttt{test\_id}-out and leave-one-\texttt{commit\_id}-out splits with identifier tokens removed, the SCOUT variants remain stable rather than collapsing, State LR reaches 0.241/0.239, and SDJ-LR reaches 0.237/0.239, respectively.  
These results show that the main dense strict causal feature is not driven by test- or commit-level shortcuts.
The explicit post-failure leakage ablation provides further evidence.
Opening the post-failure window $[-120\text{s},+30\text{s}]$ does not improve the dense LR baseline (PR-AUC 0.190), indicating that the main results are not explained by hidden post-failure evidence.
Window-length sensitivity follows the same pattern.
PR-AUC is 0.180, 0.191, and 0.199 for 30s, 60s, and 120s pre-failure windows, while extending the window to 240s does not improve further because the telemetry artifact is concentrated within the last 120 seconds.

\begin{figure}[t]
\centering
\includegraphics[width=\linewidth]{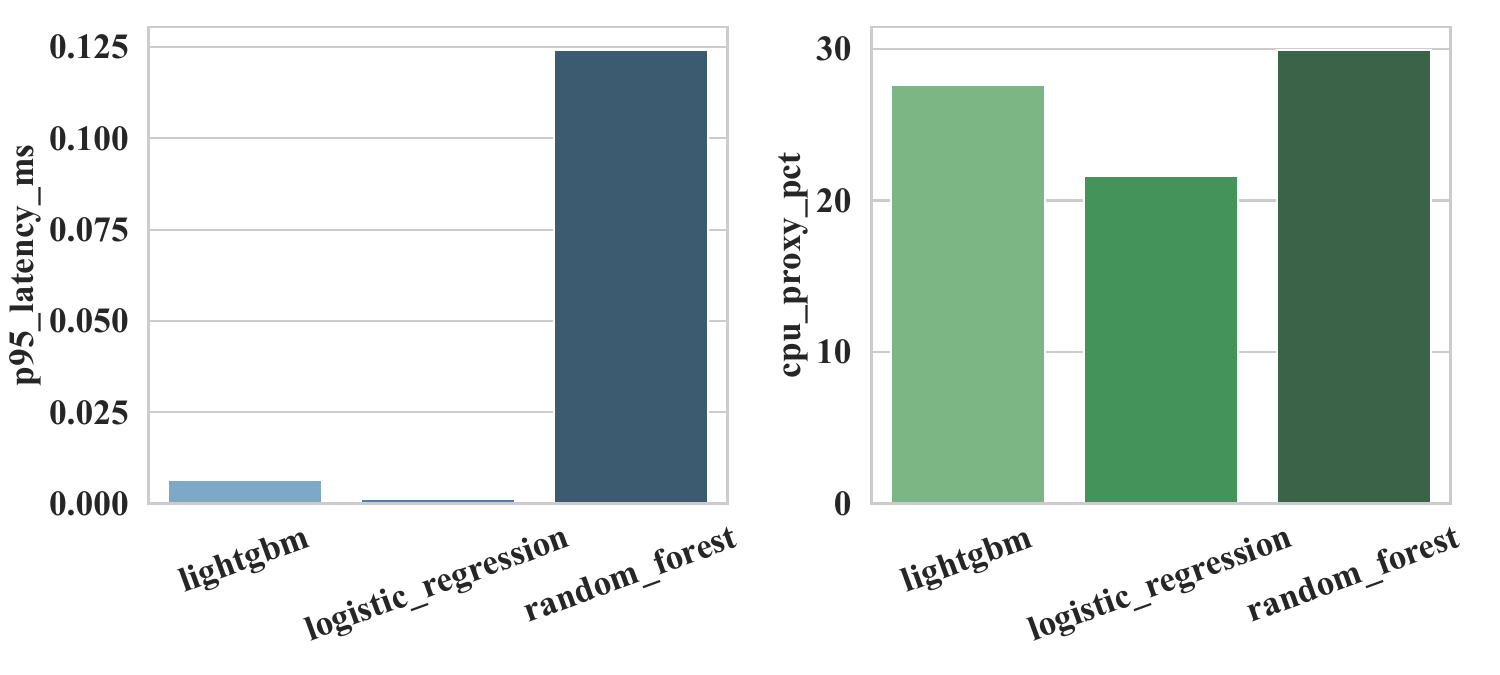}
\Description{Bar chart breaking down CPU-only latency of the triage serving path into feature extraction and model inference, highlighting millisecond-scale feasibility.}
\caption{CPU-only latency breakdown in Apple M4. We separate model-only inference from the end-to-end serving path to avoid under-reporting deployment latency.}
\label{fig:rq4_latency}
\end{figure}

Figure~\ref{fig:rq4_latency} summarizes the CPU-only latency breakdown.
On Apple M4, the end-to-end online triage path has P95 latency 1.17\,ms in the state-only pipeline, with feature extraction 0.77\,ms and inference 0.40\,ms.
For SDJ-LR, the additional sparse processing cost remains negligible relative to feature extraction, including vectorization, scaling, and inference, which together contribute only P95 0.0074\,ms.
In a \texttt{Linux/AMD64}, model inference is P95 3.0\,$\mu$s/example and the strict-causal aggregation kernel is P95 0.77\,ms/example. \
These results indicate that SCOUT is compatible with millisecond-scale CPU-only deployment as long as telemetry retrieval is not performed synchronously over remote queries.
The four Prometheus instant queries used in external validation have P95 35.9\,ms with keep-alive and 38.5\,ms cold, which suggests that production deployment should rely on local agents, cached telemetry, or recording rules rather than per-decision remote fetching.

\begin{figure}[t]
\centering
\includegraphics[width=\linewidth]{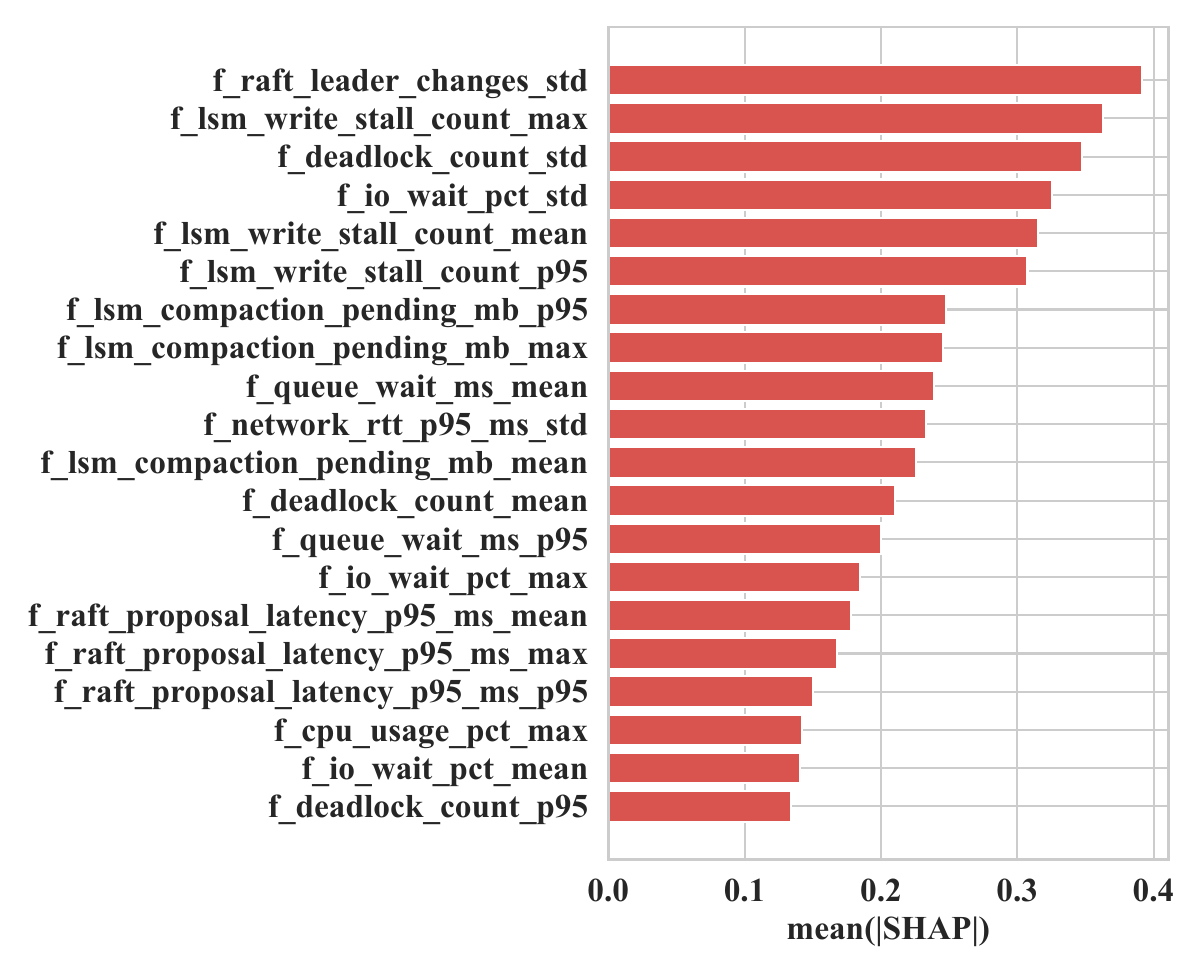}\\[2pt]
\includegraphics[width=\linewidth]{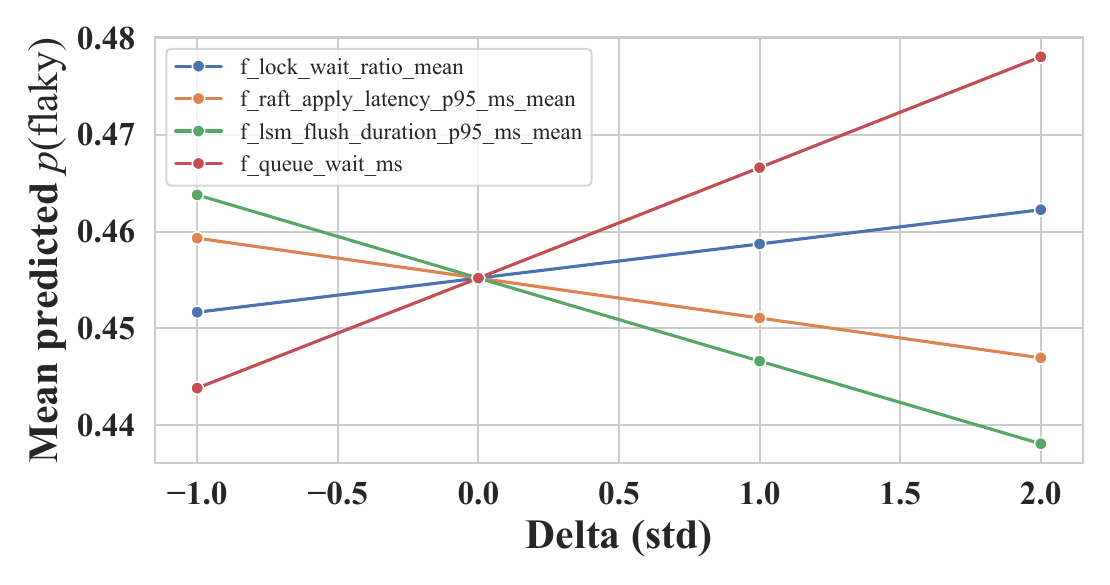}
\Description{Two-panel interpretability figure: top shows the global feature contribution ranking for the state-only logistic regression model; bottom shows a perturbation-based face-validity check of predicted flaky probability as individual features are increased.}
\caption{Top: global interpretability summary (state-only LR, strict-causal temporal split). Bottom: face-validity check via do-style perturbations on canonical state features (increase one standardized feature at a time and measure mean predicted \(p(\mathrm{flaky})\)).}
\label{fig:rq4_interpretability}
\end{figure}

Figure~\ref{fig:rq4_interpretability} shows that the learned dense strict-causal model is interpretable at both the global and local levels.
The SHAP top-20 summary highlights intuitive runtime-state drivers of flakiness, including lock-wait ratio, Raft apply latency, LSM flush duration, and queue wait.
The coefficient signs are also stable and operationally plausible.
The LSM compaction backlog ($+0.27$), queue-wait P95 ($+0.16$), and network RTT P95 ($+0.14$) increase predicted flakiness, whereas persistent severity features such as LSM write-stall count ($-0.35$), deadlock count ($-0.17$), and Raft proposal latency ($-0.12$) decrease it.
The face-validity perturbation check further confirms this behavior.
Increasing a canonical transient-state feature monotonically increases the mean predicted flaky probability, without causing unstable or counterintuitive responses.

SCOUT satisfies the key requirements of deployment practicality.
Its strict causal performance does not depend on obvious identifiers or post-failure leakage; its online path remains compatible with millisecond-scale CPU-only serving; and its state-aware logistic scoring rule is sufficiently interpretable for operational debugging and auditing.

\subsection{Generalization In Production (RQ5)}

\begin{table}[!t]
\centering
\caption{TiDB external validation on real services: PR-AUC on labeled failures under repeated 5-fold CV and cross-version transfer. Symptom baselines use only short error strings (no telemetry), and DistilBERT is a frozen offline encoder baseline. Columns shaded in light orange and marked with $\dagger$ denote SCOUT scoring variants; unshaded columns are baselines.}
\label{tab:rq5_docker_tidy}
\small
\setlength{\tabcolsep}{1pt}
\begin{tabular}{lccccc}
\toprule
Setting & \cellcolor{scoutscore}State LR$^\dagger$ & TF-IDF & DistilBERT & Dict & \cellcolor{scoutscore}SDJ-LR$^\dagger$ \\
\midrule
Repeated CV (mean) & \cellcolor{scoutscore}\textbf{0.623} & 0.561 & 0.560 & 0.561 & \cellcolor{scoutscore}0.613 \\
$v8 \rightarrow v7$ & \cellcolor{scoutscore}0.541 &0.531 & 0.531 & 0.531 & \cellcolor{scoutscore}\textbf{0.556} \\
$v7 \rightarrow v8$ & \cellcolor{scoutscore}\textbf{0.633} & 0.595 & 0.596 & 0.595 & \cellcolor{scoutscore}0.630 \\
\bottomrule
\end{tabular}
\end{table}

\begin{table}[!t]
\centering
\caption{GitHub Actions metadata-only validation (sha-group episodes; 36 repositories): fixed-threshold decision quality on labeled-valid episodes, together with a normalized IPW cost sensitivity analysis and a coarse time-saved proxy. Tokens are governed by dropping raw \texttt{branch\_} tokens and adding coarse \texttt{branchb\_} buckets and time bins. This table reports supporting calibration baselines on the governed metadata trace.}
\label{tab:rq5_github_actions}
\scalebox{0.88}{
\small
\setlength{\tabcolsep}{6pt}
\begin{tabular}{lccccc}
\toprule
Method & PR-AUC & ECE(10) & Cost@$\tau^\ast$ & IPW Cost & Net saved (h) \\
\midrule
Uncalibrated & 0.591 & 0.140 & 728.80 & 706.43 & 60.4 \\
Sigmoid      & 0.591 & 0.039 & 606.90 & 546.60 & 51.0 \\
BetaCal      & \textbf{0.591} & 0.039 & \textbf{589.45} & \textbf{541.17} & \textbf{62.5} \\
CalibTree    & 0.561 & 0.042 & 608.05 & 547.28 & 50.5 \\
Isotonic     & 0.564 & \textbf{0.030} & 608.05 & 547.28 & 50.5 \\
\bottomrule
\end{tabular}
}
\end{table}

\subsubsection{setup}
For the real-service setting, we run TiDB v7 and v8 with Prometheus telemetry and controlled fault injection, producing 1,600 runs, 341 labeled failures, and 194 positives under the rerun-based label.
Telemetry includes resource statistics per container, together with a small set of Prometheus queries, including TiDB QPS, PD TSO wait, and TiKV Raft apply and scheduler latch wait.
We evaluate PR-AUC on labeled failures under repeated cross-validation and cross-version transfer.
We compare the methods, including a telemetry-only State LR scorer, sparse symptom-based baselines on short error strings (TF-IDF LR and a dictionary matcher), a heavier frozen DistilBERT symptom baseline, and the sparse+dense SDJ-LR fusion model.

For the real-CI setting, we construct a metadata-only rerun trace from public GitHub Actions.
Features are strictly metadata-only, which logs and error strings are excluded entirely, and token governance removes raw \texttt{branch\_} tokens in favor of coarse \texttt{branchb\_} buckets and temporal bins.
Because 89.1\% of episodes are censored, results on the labeled-valid subset are subject to selection bias.
Therefore, we also report a normalized IPW cost as a best-effort sensitivity analysis.

\subsubsection{Results}

Table~\ref{tab:rq5_docker_tidy} shows that SCOUT variants remain competitive on a fully real TiDB trace.
Under repeated 5-fold cross-validation, telemetry-only State LR achieves the highest mean PR-AUC (0.623), while SDJ-LR remains close at 0.613.
Both outperform symptom-only baselines based on TF-IDF (0.561), a dictionary matcher (0.561), and frozen DistilBERT embeddings (0.560).
Under cross-version transfer, the same conclusion continues to hold.
For $v8 \rightarrow v7$, the best result is SDJ-LR (0.556), followed by State LR (0.541), both ahead of the symptom-only baselines (all 0.531).
For $v7 \rightarrow v8$, State LR achieves the best PR-AUC (0.633), with SDJ-LR close behind (0.630), while the symptom-only baselines remain around 0.595-0.596. 
These results show that the core SCOUT remains effective on a real distributed database
service.

Table~\ref{tab:rq5_github_actions} evaluates fixed-threshold decisions on the labeled-valid GitHub Actions episodes.
In this setting, raw scores are poorly calibrated.
The uncalibrated system has ECE 0.140 and cost@$\tau^\ast$ 728.80.
Post-hoc calibration substantially improves deployment-oriented action quality, reducing fixed-threshold cost to 606.90 with sigmoid scaling and to 589.45 with BetaCal, while also improving ECE to 0.039.
The normalized IPW cost shows the same relative pattern, decreasing from 706.43 to 541.17 under BetaCal. 
Although PR-AUC changes little across the calibration baselines, the decision quality at the fixed threshold improves considerably, which is consistent with RQ2.

On TiDB, lightweight strict-causal telemetry is sufficient to support competitive flaky-failure triage without heavy symptom models.
On GitHub Actions, even in a metadata-only regime with severe rerun censoring and substantial token drift, post-hoc calibration continues to improve fixed-threshold decision quality.

\section{Related Work}
\paragraph{Flaky failures in CI}
Prior work studies flaky tests and intermittent CI failures using code evolution and test history~~\cite{historychurn2023}, empirical analyses of flaky behavior at scale~\cite{luo2014empirical, chromium2023}, and symptom/log matching for just-in-time detection and categorization~\cite{jitmatch2023}. Our focus is narrower, which is triage \emph{after} a failure, under CPU-only millisecond budgets, where low-level runtime state is available, and interpretability is required. \emph{History- and churn-based approaches are} primarily useful for anticipating instability \emph{before} execution, whereas our setting reacts \emph{after} a concrete failure using strict-causal pre-failure telemetry and earlier-run history.

\paragraph{Telemetry-centric operations modeling.}
Systems and AIOps work increasingly uses multimodal logs and metrics for diagnosis and monitoring, including live forensics and database failure management~\cite{liveforensics2019,agentfm2025,logdb2025}. Our strict-causal triage is a \emph{front-end decision}, rerun versus escalate under millisecond CPU budgets. 
In this sense, it is complementary to heavier diagnosis systems and can be used to gate when such systems should be invoked.
Accordingly, we avoid heavy log parsing and GPU/LLM inference and emphasize protocol rigor under deployment constraints rather than richer representation learning.

\paragraph{Calibration under shift.}
Post-hoc calibration and calibration under covariate shift are well studied~\cite{niculescu2005, guo2017calibration,kull2017beta,pampari2020}.
Our work is closest to this line but differs in emphasizing fixed-threshold actions under temporal and cross-domain shifts in CI triage. 
Rather than evaluating calibration only as a probabilistic goodness property, we study when calibrated probabilities support a single global deployment threshold across domains and compare lightweight post hoc calibrators together with transfer-oriented extensions such as cluster-level temperature scaling~\cite{gong2021dgcalib}.

\paragraph{Cost-sensitive and decision-aware calibration.}
Cost-sensitive learning emphasizes that optimal actions depend on calibrated probabilities and explicit costs~\cite{elkan2001foundations}. Recent work further formalizes calibration as a \emph{decision} primitive and studies efficient post-hoc procedures for minimizing decision loss~\cite{gopalan2025efficientcalibdecisions}. Our decision-portable evaluation follows this line by fixing a cost-derived threshold and studying whether thresholded decisions remain stable under a shift without target-domain threshold retuning.

\paragraph{Selective prediction and risk control.}
Abstention and reject-option classification have a long history~\cite{chow1970reject}. Modern distribution-free uncertainty methods such as Venn-Abers~\cite{vennabers2012} and split conformal predictors~\cite{angelopoulos2021conformal} provide a complementary way to trade cost versus coverage, and conformal methods under shift can use reweighting or weighted scores ~\cite{tibshirani2019conformalshift,weightedconformal2024}.
We include these as lightweight, deployment-compatible baselines in our evaluation.
Operationally, they can hedge when overlap/OOV diagnostics indicate high shift or when calibration windows are tiny.

\paragraph{Finite rerun budgets and PU learning.}
Rerun-budget labeling induces one-sided censoring, where smaller budgets systematically under-observe flakiness. This connects to learning from positive and unlabeled data under selection bias~\cite{bekker2018sarem,elkan2008pu}. More broadly, it is a censored-outcome problem related to classical econometric and survival settings, including Kaplan-Meier, Tobit, and Cox-style modeling~\cite{kaplan1958,tobin1958,cox1972}. Our work differs in targeting a deployable correction that uses only information available at labeling time, namely the limited rerun outcomes, rather than requiring full time-to-event supervision.
\section{Conclusion}

In this work, we introduced \textsc{SCOUT}, a practical framework for state-aware
flaky-failure triage in distributed-database CI.
SCOUT targets the deployment setting in which a failed run must be triaged online under strict-causal, CPU-only constraints.
To support this setting, it combines strict causal feature extraction, lightweight state-aware scoring, decision-portable calibration, and rerun-budget correction.
We evaluated SCOUT on both synthetic benchmarks and real-world CI traces, and the results demonstrated its effectiveness and practical usefulness for flaky-failure triage in distributed database CI.

\bibliographystyle{ACM-Reference-Format}
\bibliography{refs}


\end{document}